\documentclass[]{spie}  

\usepackage{amsmath,amsfonts,amssymb}
\usepackage{graphicx}
\usepackage[colorlinks=true, allcolors=blue]{hyperref}
\usepackage{subfigure}
\usepackage{multirow}
\usepackage{array}
\usepackage{wasysym}
\usepackage{xcolor}
\usepackage{siunitx}
\usepackage{chemformula}

\title{Additive manufacturing in ceramics: targeting lightweight mirror applications in the visible, ultraviolet and X-ray}

\author[a*]{Carolyn Atkins}
\author[a]{Younes Chahid}
\author[a, b]{Gregory Lister}
\author[a, c]{Rhys Tuck}
\author[a]{David Isherwood}
\author[d]{Nan Yu}
\author[e]{Rongyan Sun}
\author[e]{Itsuki Noto}
\author[e]{Kazuya Yamamura}
\author[f]{Marta Civitani}
\author[f]{Gabriele Vecchi}
\author[f]{Giovanni Pareschi}
\author[g]{Simon G. Alcock}
\author[g]{Ioana-Theodora Nistea}
\author[g]{Murilo, Bazan Da Silva}

\affil[a]{UK Astronomy Technology Centre, Royal Observatory, Edinburgh, EH9 3HJ, UK}
\affil[b]{Dept of Mechanical, Aerospace \& Civil Engineering, University of Manchester, M13 9PL, UK}
\affil[c]{Dept of Mechanical, Materials \& Manu. Engineering, Uni. of Nottingham, NG7 2RD, UK}
\affil[d]{School of Engineering, University of Edinburgh, Edinburgh, EH9 3FB, UK}
\affil[e]{Research Center for Precision Engineering, Osaka University, Suita, Japan}
\affil[f]{INAF Astronomical Observatory of Brera, Via E. Bianchi 46, 23807, Merate, Italy }
\affil[g]{Diamond Light Source, Harwell Science \& Innovation Campus, OX11 0QX, UK}

\authorinfo{*Further author information:\\E-mail: carolyn.atkins@stfc.ac.uk}

\begin{document} 
\maketitle

\begin{abstract}
Additive manufacturing (AM; 3D printing), which builds a structure layer-by-layer, has clear benefits in the production of lightweight mirrors for astronomy, as it can create optimised lightweight structures and combine multiple components into one. AM aluminium mirrors have been reported that demonstrate a 44\% reduction in mass from an equivalent solid and the consolidation of nine parts into one. However, there is a limit on the micro-roughness that can be achieved using AM aluminium at $\sim$\SI{5}{\nm} RMS (root mean square; Sq), therefore, to target applications at shorter wavelengths alternative AM materials are required. New capabilities in AM ceramics, silicon carbide infiltrated with silicon (SiC + Si) and fused silica, offer the possibility to combine the design benefits of AM with a material suitable for visible, ultraviolet and X-ray applications. 

This paper will introduce the different printing methods and post-processing steps to convert AM ceramic samples into reflective mirrors. The samples are flat disks, \SI{50}{\mm} diameter and \SI{5}{\mm} in height, with three samples printed in SiC + Si and three printed in fused silica. Early results in polishing the SiC + Si material demonstrated that a micro-roughness of $\sim$\SI{2}{\nm} Sq could be achieved. To build on this study, the \SI{50}{\mm} SiC + Si samples had three different AM finishing steps to explore the best approach for abrasive lapping and polishing, the reflective surfaces achieved demonstrated micro-roughness values varied between \SI{2}{\nm} and \SI{5}{\nm} Sq for the different AM finishing steps. To date, the printed fused silica material has heritage in lens applications; however, its suitability for mirror fabrication was to be determined. Abrasive lapping and polishing was used to process the fused silica to reflective surface and an average micro-roughness of $<$\SI{1}{\nm} Sq achieved on the samples.    
\end{abstract}


\section{INTRODUCTION}
\label{sec:intro}  
Additive manufacturing (AM; 3D printing) is a method of manufacture that prints a part layer-by-layer, this method of manufacture enables significant design freedom through the direct incorporation of low mass lattices and the recreation of organic-styled structures. Using AM to create low mass (lightweight) mirrors for astronomy and Earth observation has clear benefits by leveraging the design freedom to create lattices more optimised for function (manufacture \& operation), and to incorporate mounting structures and fasteners within the mirror structure, thereby reducing part count and enabling material continuity. The majority of AM mirrors to date are made in aluminium and created using the AM method laser powder bed fusion (L-PBF), where metal powder is fused together using a laser~\cite{Atkins2021}. AM aluminium mirrors have the potential to exhibit good micro-roughness\footnote{Note - this paper will refer to micro-roughness as a description of a reflective (specular) surface, surface roughness to refer to processed surfaces (non-specular), and surface texture to refer to rough as-printed surfaces.} with $\sim$ \SIrange{4}{5}{\nm} Sq possible on the AM substrate after single point diamond turning~\cite{Atkins2018, Atkins2019a} (SPDT) and $<$ \SI{4}{\nm} Sq after additional polishing steps~\cite{Sweeney2015, Woodard17}. An alternative use of AM aluminium is to coat the AM aluminium substrate in nickel phosphorous (NiP) prior to the polishing steps to leverage the mechanical properties of nickel (increased hardness) to deliver an improved micro-roughness, with $<$ \SI{1}{\nm} Sq demonstrated~\cite{Hilpert2018, Hilpert2019}. However, although it is possible to achieve the low micro-roughness required for short wavelength applications (visible, UV and X-ray), metal mirrors do have disadvantages such as a high coefficients of thermal expansion (CTE) and a potential mismatch in the CTE in the case of aluminium + NiP, which requires careful thermal management. Therefore, there is a clear role for the use of AM ceramics, with desirable CTE and mechanical properties, to deliver low mass mirrors for short wavelength applications.  

Micro-roughness affects how photons are scattered from a surface and it is a property that is dependent on wavelength: the shorter the wavelength, the lower the micro-roughness needs to be to minimise scatter. For example, for the same percentage of scattering, a optical component would have to have a lower micro-roughness at X-ray, than at visible wavelengths. When considering optical requirements for a telescope and/or instrument, where there might be multiple optical surfaces, the error budget related to scatter is `divided' between the surfaces, thereby creating a situation where a UV and X-ray micro-roughness requirement might be similar ($\le$\SI{0.5}{\nm} Sq)~\cite{Spiga2023, Ramsey2022} if there are more surfaces in the UV system than the X-ray. Ceramics have significant heritage in short wavelength applications: \textit{Euclid}~\cite{Bougoin2017} (near infrared \& visible; SiC),  \textit{Hubble Space Telescope}~\cite{McCarthy1982} (visible \& UV; ULE, ultra low expansion glass), \textit{GALEX}~\cite{Martin2003} (UV; fused silica), and \textit{Chandra X-ray Observatory}~\cite{Weisskopf2003} (X-ray; Zerodur). Therefore, there is a clear opportunity to link the design benefits of AM with the optical heritage of ceramic materials.

There are a number of ceramics available via AM that have been explored for mirror applications to date, for example, alumina~\cite{Roulet2020a}, cordierite~\cite{Roulet2020a} and SiC~\cite{Goodman2019, Goodman2021a, Goodman2021b, HORVATH2020}. The focus of this paper will to add to the on-going research relating to SiC and to introduce opportunities with fused silica. Previous research into SiC has looked at two AM methods: binder jetting in SiC followed by a chemical vapour infiltration of SiC; and fused deposition modelling, termed direct ink writing, using a SiC `nanopaste'. Binder jetting is an AM method that uses a powdered form of the required material and prints the part through successive `glue' layers. The glue, an organic binder, is removed from the printed part in a secondary step using heat, tertiary steps are required to convert the printed material into a solid, fully dense, part. In \textit{Horvath, et al. (2020)}~\cite{HORVATH2020}, the binder jetted SiC had a tertiary step of chemical vapour infiltration in SiC. In this initial experiment, the micro-roughness achieved on the \SI{50}{\mm} diameter work piece was \SI{104}{\nm} Sq. In the alternative approach, the SiC nanopaste was extruded from a nozzle, with secondary steps required to finish the SiC composite material~\cite{Goodman2021b}. Several mirror prototype substrates have been demonstrated via this approach~\cite{Goodman2021a, Goodman2021b}; however, only one prototype has a recorded micro-roughness associated with a specular surface (\SI{2}{\nm} Sq) and was described as ``too soft to be dense silicon carbide''~\cite{Goodman2019}.

In contrast, there is no documented research in AM fused silica for mirror applications; however, fused silica~\cite{Kotz2017} and fused quartz~\cite{Luo2016} have both either demonstrated or suggested applications for broader optical components: lenses and filters. In the paper by \textit{Kotz, et al. (2017)}~\cite{Kotz2017}, the AM method vat photopolymerisation is used to create the printed fused silica. In this method a liquid resin is cured with UV light in layers to print the part, the liquid resin contains fused silica particles within an organic resin. After printing the resin is removed via a heat treatment and the fused silica formed in a final sintering phase.

The aim of this paper is to explore the potential of AM SiC and AM fused silica to deliver reflective surfaces (low micro-roughness) for short wavelength applications (visible, UV and X-ray). The objective for AM SiC is to build on an earlier study (the OPTICON Prototype; Section~\ref{sec:IAC}) using binder jetting and to trial different post-print finishing steps to understand how these may affect the resultant reflective surface after polishing. The objective for AM fused silica is provide a foundational knowledge in the ability of AM fused silica, printed via vat photopolymerisation, to be lapped and polished to generate a reflective surface. The paper commences with an introduction to an initial study using SiC to create a low mass mirror substrate (Section~\ref{sec:IAC}), which provides the foundation of the SiC objective. The samples used in the SiC and fused silica objectives are identical, therefore the SiC and fused silica samples will run in parallel through the Method (Section~\ref{sec:method}) and the Metrology Results (Section~\ref{sec:metrol}). The paper concludes with a summary and future works (Section~\ref{sec:future}).  

\section{Proof of concept: SiC + Si}
\label{sec:IAC}
The OPTICON prototype originated from a series of test prints exploring the use of silicon carbide infiltrated with silicon (SiC + Si) as a potential material for mirror fabrication, which were conducted as part of work package 5 (WP5) of the EU Horizon 2020 funded project OPTICON (Optical Infrared Coordination Network for Astronomy; grant agreement \#730890). WP5, termed `Additive Astronomy Integrated-component Manufacturing' (A2IM), explored the disruptive potential of AM within astronomical instrumentation and was a collaborative effort between seven European partners. A key deliverable of WP5 was the OPTICON A2IM Cookbook~\cite{Atkins2022, Atkins2021}, within this document the OPTICON prototype is described as Case Study \#5.    

\subsection{Prototype design}
\begin{figure}
    \centering
    \includegraphics[width = 0.95\textwidth]{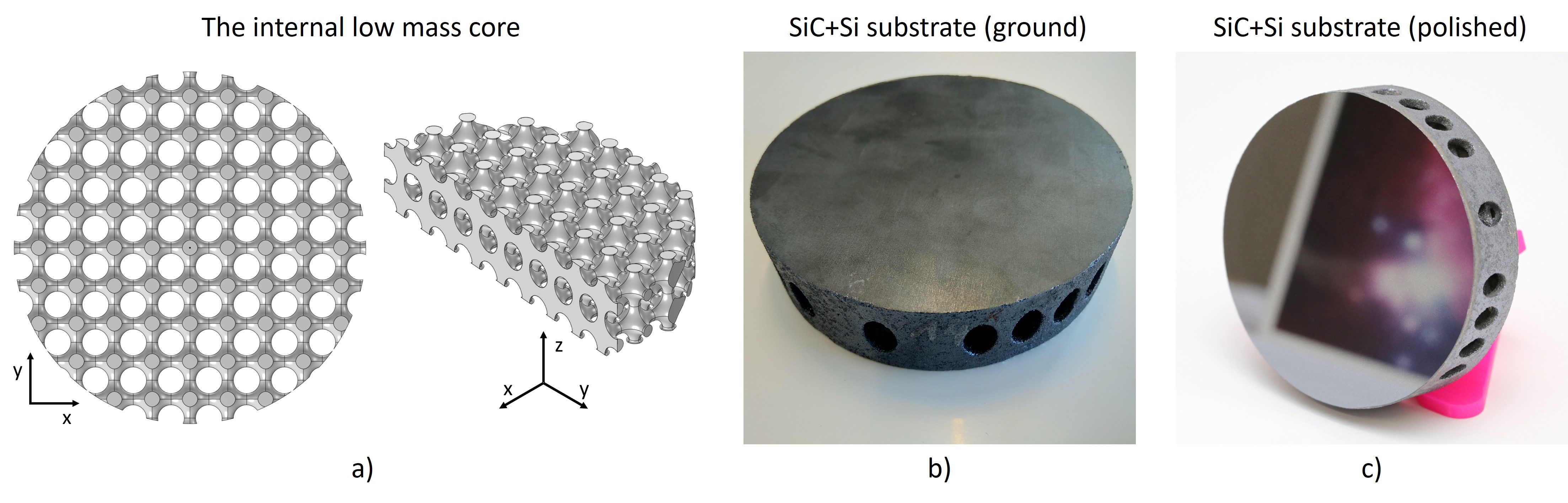}
    \caption{The OPTICON prototype: \textit{a)} a CAD representation of the internal lightweight structure; \textit{b)} the ground surface of the prototype (photo credit, L. T. G. van de Vorst); and \textit{c)} the polished optical surface.}
    \label{fig:IACmir}
\end{figure}

The prototype was designed through a collaboration of the Instituto de Astrof\'{i}sica de Canarias (Spain) and TNO-Eindhoven (The Netherlands). The prototype is optically flat with a diameter of \SI{100}{\mm} and a height of \SI{26}{\mm}, with an internal low mass core created using a triply periodic minimal surface (TPMS) Schwartz P lattice with a cubic unit cell size of \SI{10}{\mm}. The lightweight mirror design is nominally a sandwich structure, which has a \SI{3}{\mm} thick face plate, base plate and side wall. The internal lattice reduces the mass of the prototype to $\sim$47\% of an equivalent solid. Figure~\ref{fig:IACmir} \textit{a} presents the internal TPMS Schwartz P lattice core.    

\subsection{Prototype manufacture}\label{subsec:IACmanu}
The prototype was printed in SiC + Si; trade name SICAPRINT + Si~\cite{SGLCarbon2024} (SGL Carbon; Germany). In this method the SiC is printed via binder jetting, where a bed of SiC powder is fused together using a binding agent to create the 3D structure. The binding agent is then removed from the part and the porous SiC 3D structure is infiltrated with liquid silicon. The surface texture of SiC + Si, measured on sample cubes with dimensions \SI{40}{\mm} $\times$ \SI{40}{\mm} $\times$ \SI{27.5}{\mm} using a Intra-Touch contact profilometer (Taylor Hobson, UK) with a diamond tip stylus, is $\sim$\SI{9}{\micro\metre} using a \SI{0.8}{\mm} cut-off ($\lambda_{c}$) filter and $\sim$\SI{11}{\micro\metre} for a \SI{2.5}{\mm} cut-off filter.

\begin{figure}
    \centering
    \includegraphics[width = 0.95\textwidth]{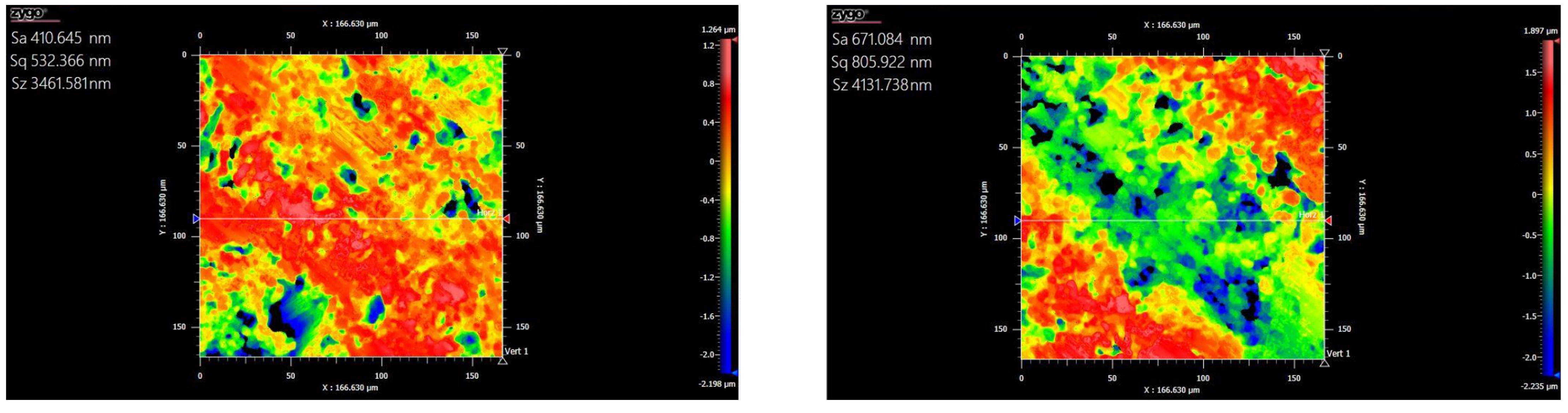}
    \caption{Surface texture measurements taken at Osaka University of the ground OPTICON prototype prior to lapping - field of view \SI{167}{\micro\metre} $\times$ \SI{167}{\micro\metre}.}
    \label{fig:IAC_ground}
\end{figure}

The prototype was ground at the supplier and the resultant surface can be seen in Figure~\ref{fig:IACmir} \textit{b}. The roughness was measured using Zygo NewView (ZYGO, USA) white light interferometer by Osaka University prior to lapping\footnote{Note - in this paper lapping defined as the process to remove significant material to create the nominal optical form, whereas polishing is defined as the process to create the specular surface.} (Figure~\ref{fig:IAC_ground}), the average Sq and standard deviation from three measurements are \SI{625}{\nm} and \SI{128}{\nm} to three significant figures respectively. Through successive lapping cycles and polishing cycles, a highly reflective, specular, surface was achieved. 

\subsection{Prototype micro-roughness}\label{subsec:IACmet}

\begin{figure}
    \centering
    \includegraphics[width = 0.95\textwidth]{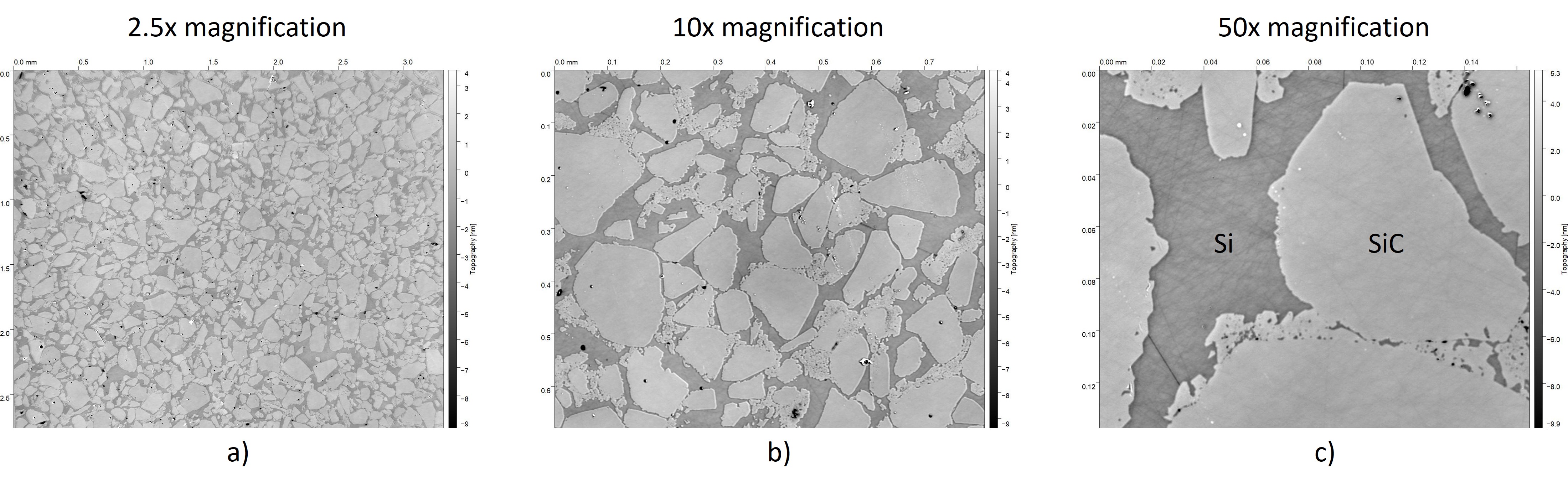}
    \caption{Example micro-roughness measurements taken at the Diamond Light Source highlighting the SiC and Si matrix: \textit{a)} 2.5x magnification, \textit{b)} x10 magnification, and \textit{c)} x50 magnification.}
    \label{fig:IAC_GTX}
\end{figure}

\begin{figure}
    \centering
    \includegraphics[width = 0.95\textwidth]{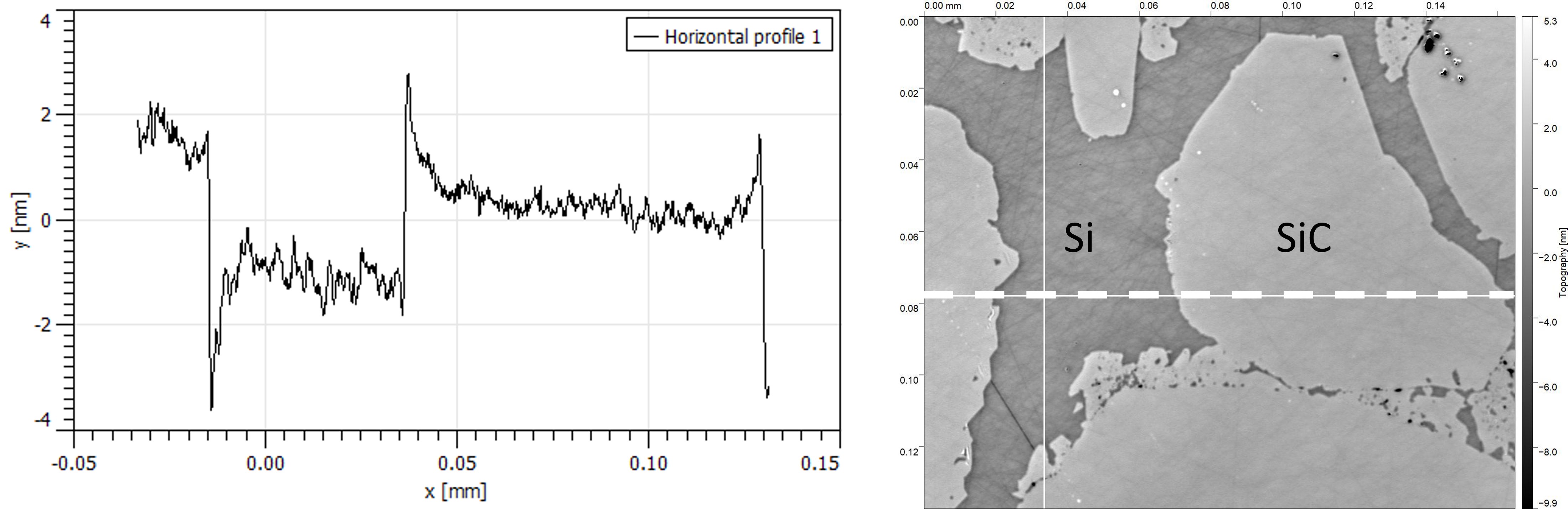}
    \caption{The observed steps between the SiC and the Si materials: \textit{left} a cross-section transitioning from SiC $\rightarrow$ Si $\rightarrow$ SiC; \textit{right} the location (dashed line) in the interferogram where the profile was taken.}
    \label{fig:IAC_GTXprofile}    
\end{figure}

Independent micro-roughness measurements were undertaken at the Diamond Light Source using a Contour GTX white light micro-interferometer (Bruker, USA) at three magnifications: x2.5, x10, and x50. Four measurements were sampled across the surface, one in centre (0,0) and three \SI{25}{\mm} offset from the centre at (-25,0), (25, 0), and (0, -25); data was not recorded for (0, 25). Table~\ref{tab:Bruker} and Figure~\ref{fig:IAC_GTX} present the numerical micro-roughness parameters (average Sa; Sq; and peak to valley, Sz) and interferograms for the three magnifications. The Sq values range from $\sim$\SI{2}{\nm} to $\sim$\SI{3}{\nm} over the four measurement areas and three magnifications. Evaluation of cross-sections from the x50 data demonstrates the presence of a step between the Si and the SiC (Figure~\ref{fig:IAC_GTXprofile}), which is a result of the mechanically harder \ch{SiC} having a lower polishing removal rate than \ch{Si}. The observed peak and trough at the transitions between materials are considered a measurement artefact of the white light micro-interferometer, as such, the exact magnitude of the step cannot be extracted from the cross-section - alternative metrology techniques are required.  

{\renewcommand{\arraystretch}{1.25}
\begin{table}[h]
\centering
\caption{Micro-roughness measurements taken at the Diamond Light Source; measurement locations are in [mm] from the centre (0, 0).}
\begin{tabular}{|l||l|l|l||l|l|l||l|l|l|}
\hline
& \multicolumn{3}{|c||}{x2.5 magnification}	& \multicolumn{3}{|c||}{x10 magnification} & \multicolumn{3}{|c|}{x50 magnification} \\
& \multicolumn{3}{|c||}{[\SI{3.31}{\mm}$\times$\SI{2.75}{\mm}]}	& \multicolumn{3}{|c||}{[\SI{814}{\micro\metre}$\times$\SI{678}{\micro\metre}]} & \multicolumn{3}{|c|}{\SI{137}{\micro\metre}$\times$\SI{165}{\micro\metre}]} \\
\hline
Meas.\# &	Sa [nm] & Sq [nm] & Sz [nm] & Sa [nm] & Sq [nm] & Sz [nm] & Sa [nm] & Sq [nm]& Sz [nm] \\
\hline
(-25,0) & 0.84 & 1.37 & 282 & 1.06 & 2.12 & 371 & 1.04 & 2.07 & 557 \\
(0, 25) & - & - & - & - & - & - & - & - & - \\	
(25, 0) & 0.96 & 1.33 & 257 & 1.35 & 2.27 & 175 & 1.92 & 3.14 & 179 \\
(0, -25) & 1.12 & 1.91 & 280 & 1.45 & 2.48 & 175 & 1.75 & 4.96 & 443 \\
(0, 0) & 0.77 & 2.82 & 437 & 1.10 & 5.39 & 730 & 1.07 & 2.37 & 192 \\
\hline
\multicolumn{10}{|l|}{}\\
\hline
\textbf{Ave.} & \textbf{0.92} & \textbf{1.86} & \textbf{314} & \textbf{1.24} & \textbf{3.06} & \textbf{363} & \textbf{1.44} & \textbf{3.13} & \textbf{343} \\
Std & 0.13 & 0.60 & 72 & 0.16 & 1.35 & 227 & 0.39 & 1.12 & 162 \\
\hline
\multicolumn{10}{l}{}\\
\end{tabular}
\label{tab:Bruker}
\end{table}}

\subsection{Prototype conclusion}
The micro-roughness results highlighted the potential of the SiC + Si material for short wavelength applications. The observed step between the \ch{Si} and \ch{SiC} materials is considered the dominant contributor to the micro-roughness, eliminating this step would, it is hoped, shift the micro-roughness to smaller values leading to applications in the UV and X-ray.

Through discussions with the supplier, alternative finishing processes are available for the SICAPRINT+Si matrix. The OPTICON prototype was created using a 1x infiltration with liquid silicon that gave a material composition of 70\% \ch{SiC}, 25\% \ch{Si}, and 5\% \ch{C}. However, there are options for a x2 infiltration (84\% \ch{SiC}, 15\% \ch{Si}, and 1\% \ch{C}) and a SiC coating created using chemical vapour deposition (CVD). Therefore, to build upon this encouraging result, small samples incorporating the alternative finishing processes were proposed and this study was run in parallel with the AM fused silica study.  

\section{Method}
\label{sec:method}  

\subsection{Experimental procedure}
A flat disk was designed to simplify the investigation into the quality of the AM SiC + Si and AM fused silica after lapping and polishing. The disk was \SI{50}{\milli\metre} in diameter and \SI{5}{\milli\metre} in height. A \SI{1}{\milli\metre} chamfer was added to the optical surface and fiducials were added every \ang{90} around the circumference. Three disks of both AM ceramics were printed to assess repeatability in fused silica and to explore alternative finishing processes in SiC + Si. The fused silica and SiC + Si samples were processed at INAF Astronomical Observatory of Brera and Osaka University respectively.

\subsection{Fused silica}\label{subsec:fusedSimethod}
In this study, the three fused silica samples were printed commercially by Glassomer GmbH (Germany). 
\subsubsection{Fused silica AM method}
\begin{figure}
    \centering
    \includegraphics[width = 0.95\textwidth]{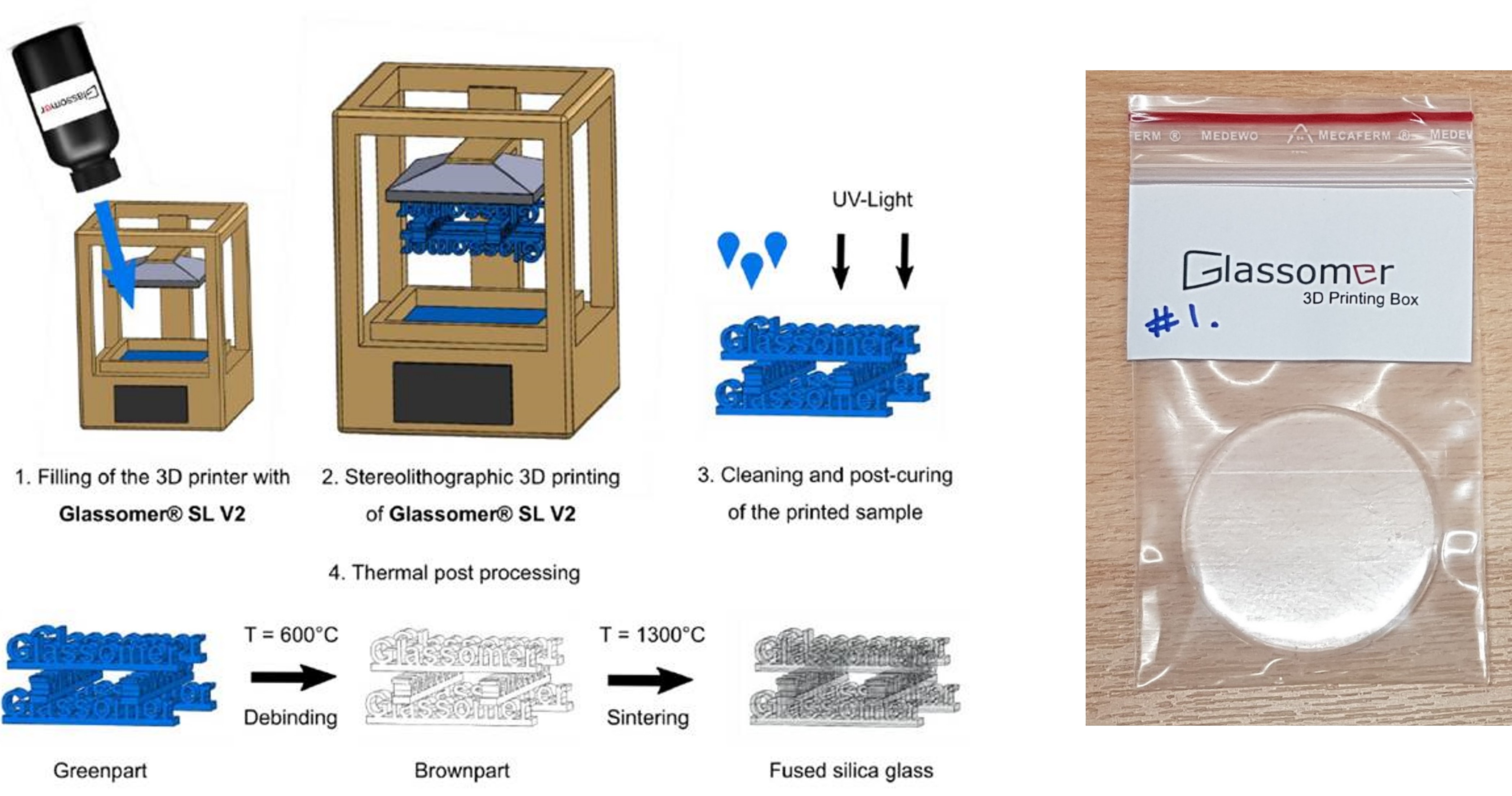}
    \caption{\textit{Left} the Glassomer AM processing steps from printing to fused silica glass (image copyright Glassomer\textsuperscript{\textcopyright}, used with permission); and \textit{right} an example of a printed fused silica disk.}
    \label{fig:glassomer}
\end{figure}

The fused silica was printed using vat photopolymerisation, in this application, the resin is a combination of fused silica particles and an organic binder~\cite{Glassomer2024}. Digital Light Processing (DLP), a vat photopolymerisation technology, projects a UV cross section of the design to solidify the organic binder in layers to create the part. Next, the printed 3D structure is heated to \SI{600}{\degreeCelsius} to remove the organic binder (debinding) and the remaining ceramic is sintered at \SI{1300}{\degreeCelsius} to provide a dense fused silica material. The process steps are presented in Figure~\ref{fig:glassomer}; in the production of the samples for this study an Asiga Pro DLP printer (Asiga, Australia) was used, delivering a layer height of \SI{50}{\micro\metre} and a spatial resolution accuracy of \SI{80}{\micro\metre}.   

\subsubsection{Fused silica material \& post-print characteristics}
Table~\ref{tab:MatProp_FS} presents the listed optical and thermal characteristics of the printed fused silica. 

{\renewcommand{\arraystretch}{1.25}
\begin{table}[h]
\centering
\caption{Optical and thermal characteristics of printed fused silica~\cite{Glassomer2024}}
\begin{tabular}{|l|l|}
\hline
Parameter & Value \\
\hline
UV transmission at \SI{200}{\nano\metre} & 85\%\\
VIS transmission at \SIrange{300}{1000}{\nano\metre} & $>$92\%\\
IR transmission at \SIrange{1000}{3400}{\nano\metre} & $>$90\%\\
Coefficient of thermal expansion & \SI{0.52e-6}{1\per\kelvin} \\
Density & \SI{2.2}{} $\pm$ \SI{0.022}{\gram\per\cubic\cm} \\
\hline
\end{tabular}
\label{tab:MatProp_FS}
\end{table}}

After print, qualitatively each sample was semi-transparent with a smooth, but textured, surface. Illuminating the samples using white light highlighted the presence of inclusions within the material. Figure~\ref{fig:meth_FS_profil} \textit{left} presents the surface quality of an as-received sample prior to optical fabrication.

\subsubsection{Fused silica optical fabrication}
\begin{figure}
    \centering
    \includegraphics[width = 0.95\textwidth]{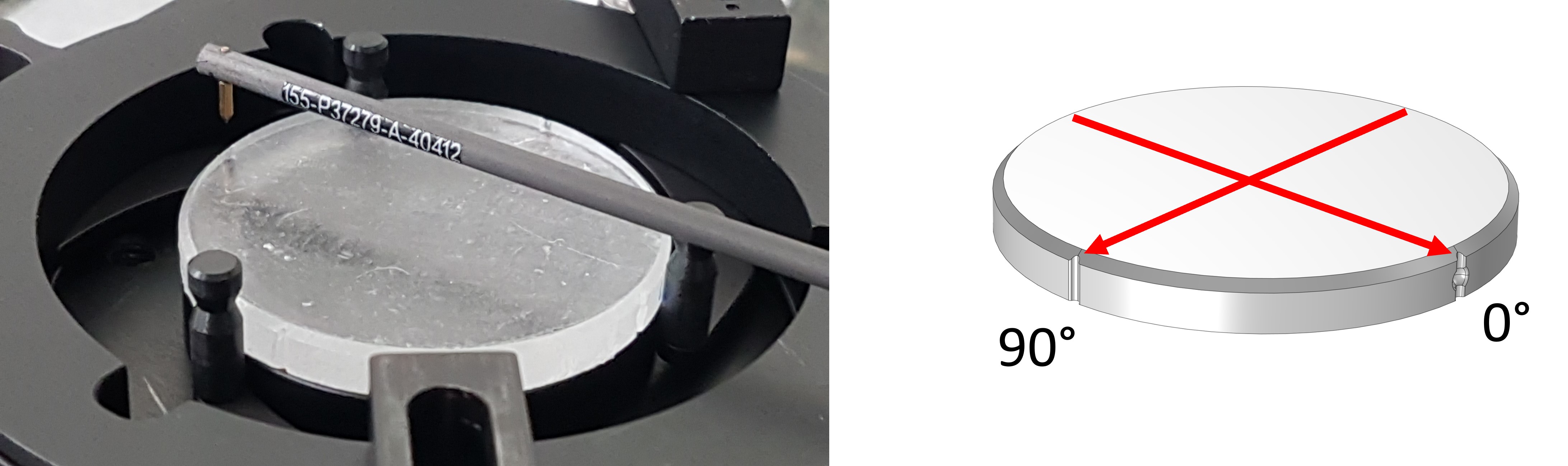}
    \caption{\textit{Left} measurement of an as-printed fused silica sample using the Intra-touch, \textit{right} the measurement strategy of the profilometry with respect to the fiducials.}
    \label{fig:meth_FS_profil}
\end{figure}

\begin{figure}
    \centering
    \includegraphics[width = 0.95\textwidth]{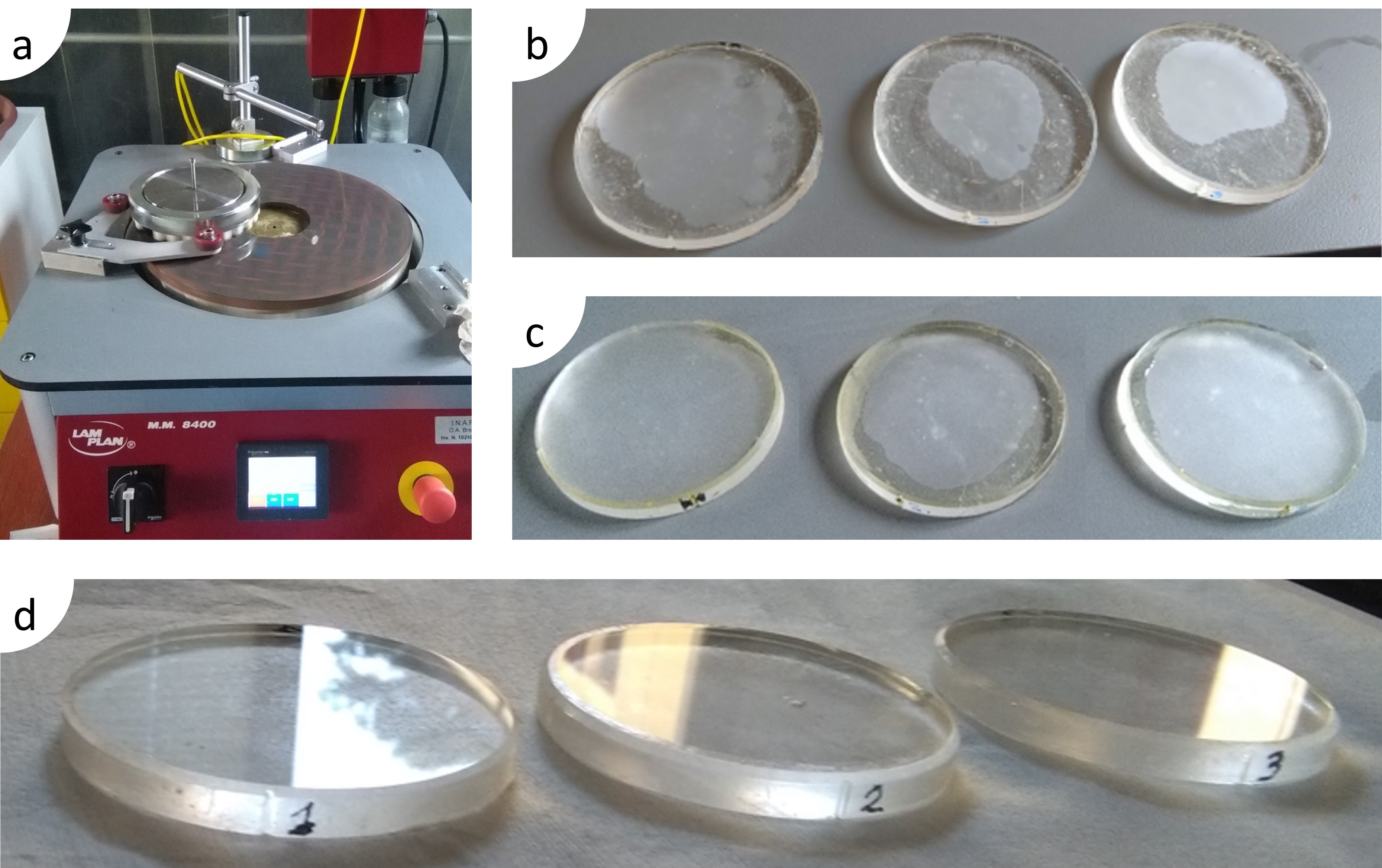}
    \caption{Lapping and polishing at INAF Brera: \textit{(a)} lapping \& polishing conducted on a M.M. 8400 lapper/polisher; \textit{(b, c)} Glassomer samples during lapping; and \textit{(d)} Glassomer samples with a specular surface after $\sim$1h of polishing.}
    \label{fig:Brera_lapping}
\end{figure}

The three identical samples were first measured across the full optical aperture using a Intra-Touch contact profilometer to assess the initial form error; two measurements were taken per sample, \ang{0} and \ang{90} (Figure~\ref{fig:meth_FS_profil}). The initial form error and surface texture was removed using iterations of lapping using a M.M. 8400 lapper/polisher (LAM PLAN, France; Figure~\ref{fig:Brera_lapping} \textit{a}), as demonstrated in Figure~\ref{fig:Brera_lapping} \textit{b} and \textit{c} with the increasing area of the diffuse surface. After each lapping cycle the surface roughness of the sample was assessed using a SurfTest SJ-210 portable contact profiler (Mitutuyo, Japan) at 10 locations uniformly sampling the optical aperture. Once uniform surface roughness was achieved across the full aperture, the transition was made to polishing, noted by a change of abrasive compound, the surface finish after the first hour of polishing is shown in Figure~\ref{fig:Brera_lapping} \textit{d}. The micro-roughness of the specular surfaces was assessed using a MicroFinish Topographer (Optical Perspectives Group, USA) to quantify the improvement of different polishing strategies to minimise the micro-roughness; the polishing ceased when no further improvements in micro-roughness were considered possible using the M.M. 8400. The surface form error achieved in the final polishing cycle on each sample was recorded using a Zygo GPI-XP interferometer (ZYGO, USA). 

\subsection{Silicon carbide infiltrated with silicon}\label{subsec:SiCmethod}
The SiC + Si samples were printed commercially at SGL Carbon; to explore the different finishing processes and the resultant affect on optical fabrication, the three samples had the following compositions:

\begin{enumerate}
    \item SICAPRINT + Si-10 (70\% SiC, 25\% Si \& 5\% C)
    \item SICAPRINT + Si-200 (84\% SiC, 15\% Si \& 1\% C)
    \item SICAPRINT + Si-10 plus $\sim$\SI{100}{\micro\metre} CVD SiC
\end{enumerate}

Sample 1 (70\% SiC) was a repeat of the of the OPTICON prototype from Section~\ref{sec:IAC}; Sample 2 (85\% SiC) explores the higher percentage of SiC, which might minimise the frequency of steps between the SiC and Si materials; and Sample 3 (CVD SiC) uses a $\sim$\SI{100}{\micro\metre} CVD coating with the objective of providing a uniform SiC layer in which to generate the optical surface. 

\subsubsection{SiC + Si AM process}
A summary of the AM process, binder jetting, was provided in Section~\ref{subsec:IACmanu}. In the final production phase, all three samples were ground as a final stage to create a flat surface from which to commence lapping. 

\subsubsection{SiC + Si material \& as-printed characteristics}

Table~\ref{tab:MatProp_SiCSi} provides a subset of material properties provided by the supplier~\cite{SGLCarbon2024}. The Si-100 variant was not used in this study; however, it has been included for completeness. 

{\renewcommand{\arraystretch}{1.25}
\begin{table}[h]
\centering
\caption{Mechanical properties of SICAPRINT + Si}
\begin{tabular}{|l|l|l|l|}
\hline
Parameter & Si-10 & Si-100 & Si-200 \\
\hline
Density [\SI{}{\gram\per\cubic\cm}] & 2.8 & 2.9 & 3.0 \\
Open porosity [\%] & $<$0.1 & $<$0.1 & $<$0.1 \\
Young's Modulus [\SI{}{\giga\pascal}] & 270 & 320 & 360 \\
Thermal conductivity [\SI{}{\watt\per\milli\kelvin}] & 120 & 150 & 170 \\
\hline
\end{tabular}
\label{tab:MatProp_SiCSi}
\end{table}}

\begin{figure}
    \centering
    \includegraphics[width = 0.95\textwidth]{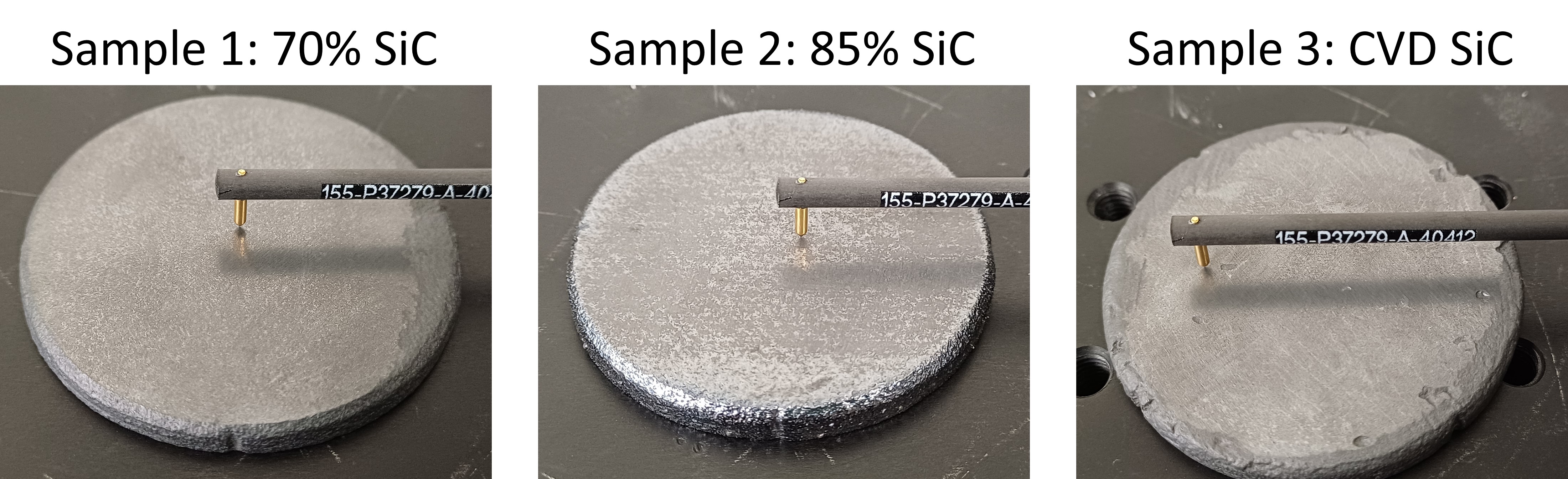}
    \caption{The surface texture of the as-printed SiC samples.}
    \label{fig:SiCsamples}
\end{figure}

Qualitatively, each of the samples exhibited a different surface texture after print, as shown in Figure~\ref{fig:SiCsamples}. Samples 1 and 3 were dull in appearance, whereas Sample 2 (85\% SiC) was more reflective. Defects from the print process were observed on Samples 1 and 2 and resulted in a non-flat surface. Due to the visible depressions observed on Sample 3 (CVD SiC), a fourth sample was created that ground the SiC + Si prior to CVD SiC; this sample was interchanged for Sample 3 and the original Sample 3 disregarded. 

\subsubsection{SiC + Si optical fabrication}
Prior to lapping, the surface roughness of the ground samples was measured using the Intra-Touch contact profilometer with a diamond tip stylus and using a filter cut-off ($\lambda_{c}$) of \SI{0.8}{\mm} and an evaluation length ($l_{n}$) of \SI{4}{\mm}. The Rq (RMS) and Rz (PV) values for the different samples were similar, with an average Rq value of \SI{0.63}{\micro\metre} and a standard deviation of \SI{0.11}{\micro\metre} and an average Rz value of \SI{3.62}{\micro\metre} with a standard deviation of \SI{0.56}{\micro\metre}. The average Rq and standard deviation are consistent with the surface roughness analysis from the OPTICON prototype (Section~\ref{subsec:IACmanu} \& Figure~\ref{fig:IAC_ground}).   

\begin{figure}
    \centering
    \includegraphics[width = 0.95\textwidth]{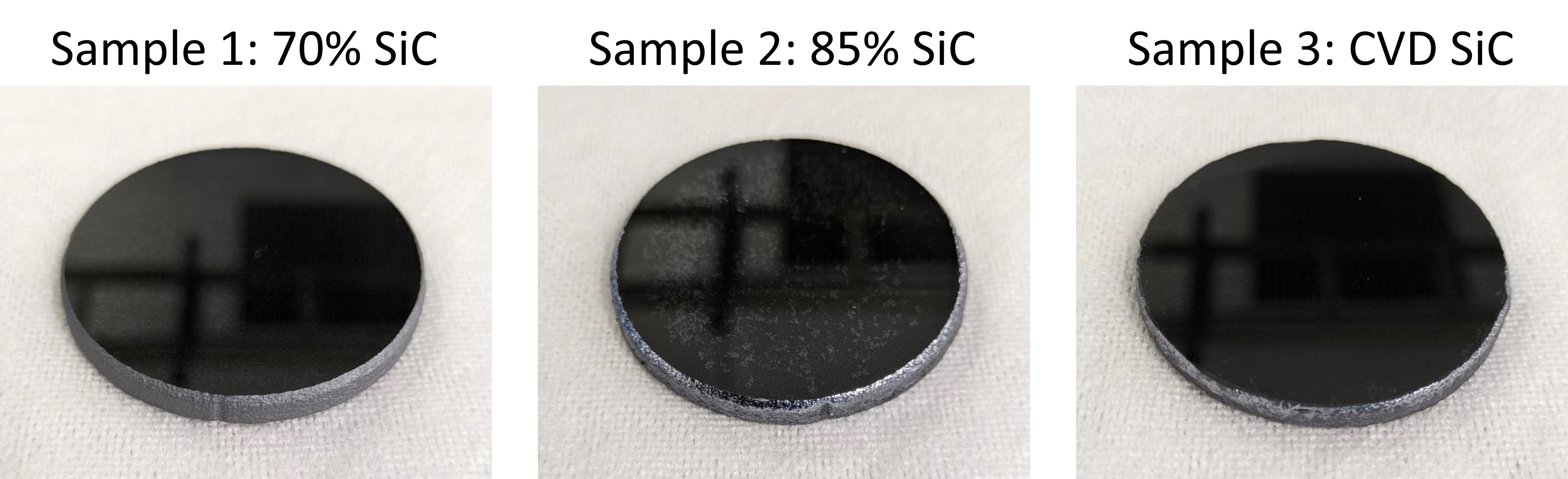}
    \caption{The three SiC + Si samples after the final polishing cycle.}
    \label{fig:SiCpolish}
\end{figure}

Lapping and polishing of the samples was undertaken at Osaka University using an Engis lapping and polishing machine (Engis, USA) and the quality of the reflective surfaces was assessed using a Zygo NewView with a square field of view of \SI{336}{\micro\metre}. Given the composite material of the samples, energy dispersive X-ray spectroscopy (EDX) was performed to evaluate elemental composition. The optical surfaces from the final cycle of polishing are shown in Figure~\ref{fig:SiCpolish}, as observed, a specular surface has been achieved on each sample.

\section{METROLOGY RESULTS}
\label{sec:metrol}

\subsection{Fused silica}
\subsubsection{As-printed}
\begin{figure}[h]
    \centering
    \includegraphics[width = 0.95\textwidth]{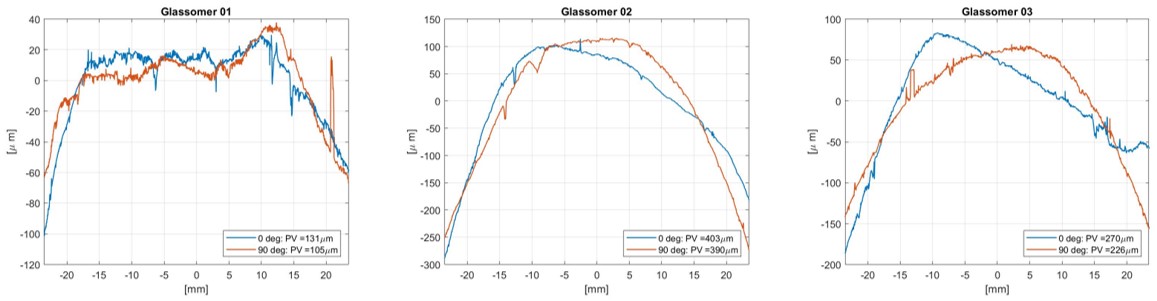}
    \caption{Contact profilometry of each fused silica sample in two orthogonal directions highlighting the as-printed surface form error.}
    \label{fig:FSmet:profil}
\end{figure}

Figure~\ref{fig:FSmet:profil} presents the contact profilometry measured at \ang{0} and \ang{90} orientations; only tilt has been removed from the data. The convex form of the profiles exhibited a peak to valley (Rz) ranging from $\sim$\SI{120}{\micro\metre} (Glassomer \#1) to $\sim$\SI{400}{\micro\metre} (Glassomer \#2). The RMS surface roughness (Rq) as measured on the SurfTest SJ-210, averaging the three samples, was $\sim$\SI{1.3}{\micro\metre}. These surface parameters represented the starting point for the lapping and polishing cycles.  

\subsubsection{Surface roughness evolution: lapping to polishing}

\begin{figure}[h]
    \centering
    \includegraphics[width = 0.95\textwidth]{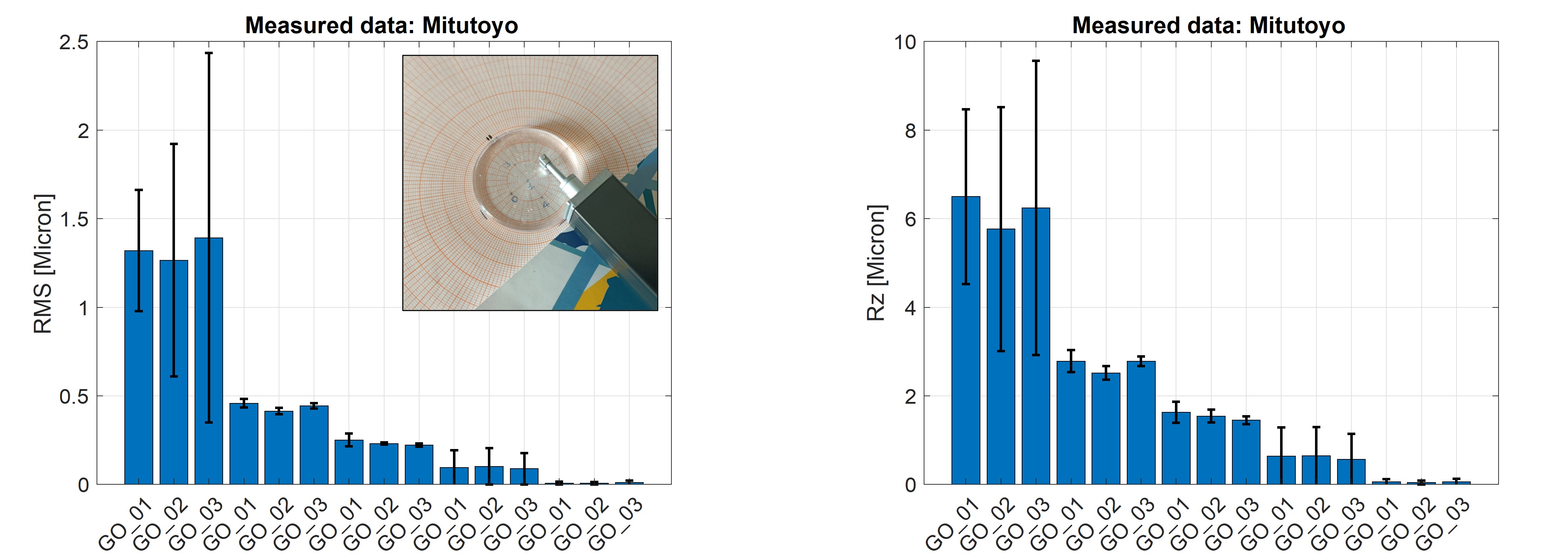}
    \caption{Surface roughness values measured using the SurfTest SJ-210: \textit{left} the RMS values (Rq), and \textit{right} the PV (Rz). The error bars represent the standard deviation of the 10 measurements per sample and the 15 data points per plot highlight five stages in lapping.}
    \label{fig:FSmet:mitutuyo}
\end{figure}

Figure~\ref{fig:FSmet:mitutuyo} presents the evolution of the surface roughness from the as-printed state to the first polishing cycle as measured on the SurfTest SJ-210. In the figure, each bar represents the average of the 10 datasets and the error bars represent the standard deviation. The final three measurements represent the first polishing cycle and highlight the significant improvement achieved after this cycle - at this stage, the SJ-210 reached the limit for accurate measurements.

\subsubsection{As-polished}
\begin{figure}[h]
    \centering
    \includegraphics[width = 0.95\textwidth]{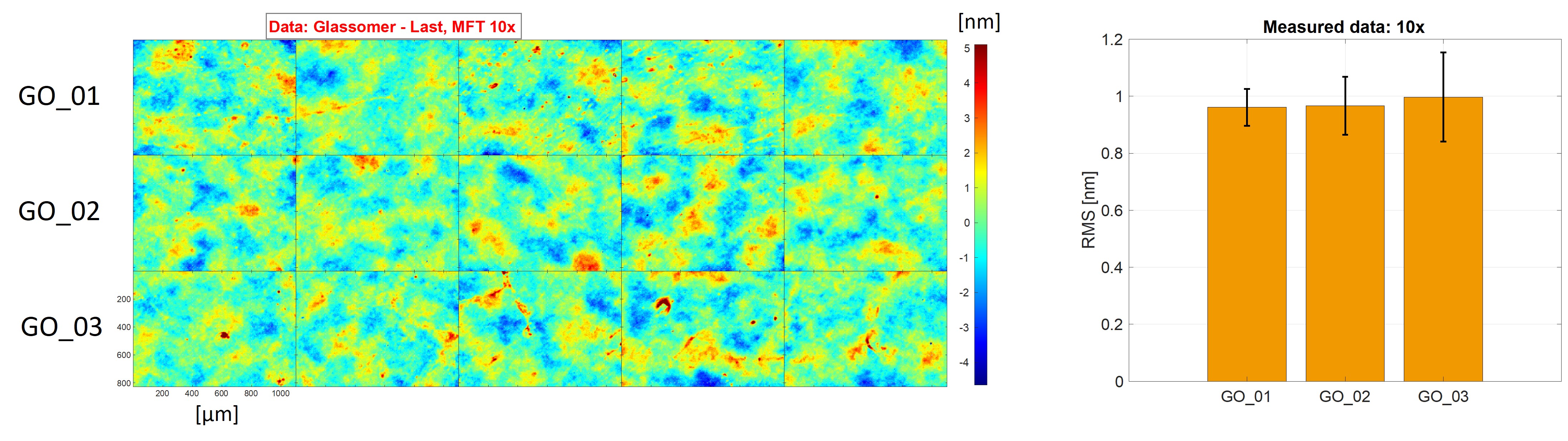}
    \caption{The surface roughness measurements of the fused silica samples: \textit{left} the five interferograms (rows) per sample (columns); \textit{right} the averaged Sq values per sample with the associated standard deviation.}
    \label{fig:FSmet:SqLast}
\end{figure}

After several cycles of polishing, the average Sq of the samples reached $<$ \SI{1}{\nm} and the polishing iterations ceased. Figure~\ref{fig:FSmet:SqLast} presents the micro-roughness data measured from the MFT on the final polishing cycle, five locations were measured on each sample and are represented as rows within the $3 \times 5$ matrix in the figure. The average Sq value, of the five measurements per sample, is represented in the bar chart with the corresponding standard deviation. The figure highlights that the material has a uniform micro-roughness and there are no localised defects which are commonly seen in AM substrates. However, micro-bubbles ($\sim$\SI{10}{\micro\metre} diameter) were suspected within some interferograms and they are defect of the AM print method. A comparison of the as-printed roughness measurements with the final polishing cycle is presented in Table~\ref{tab:FS_met:summary}. 

\begin{figure}[h]
    \centering
    \includegraphics[width = 0.95\textwidth]{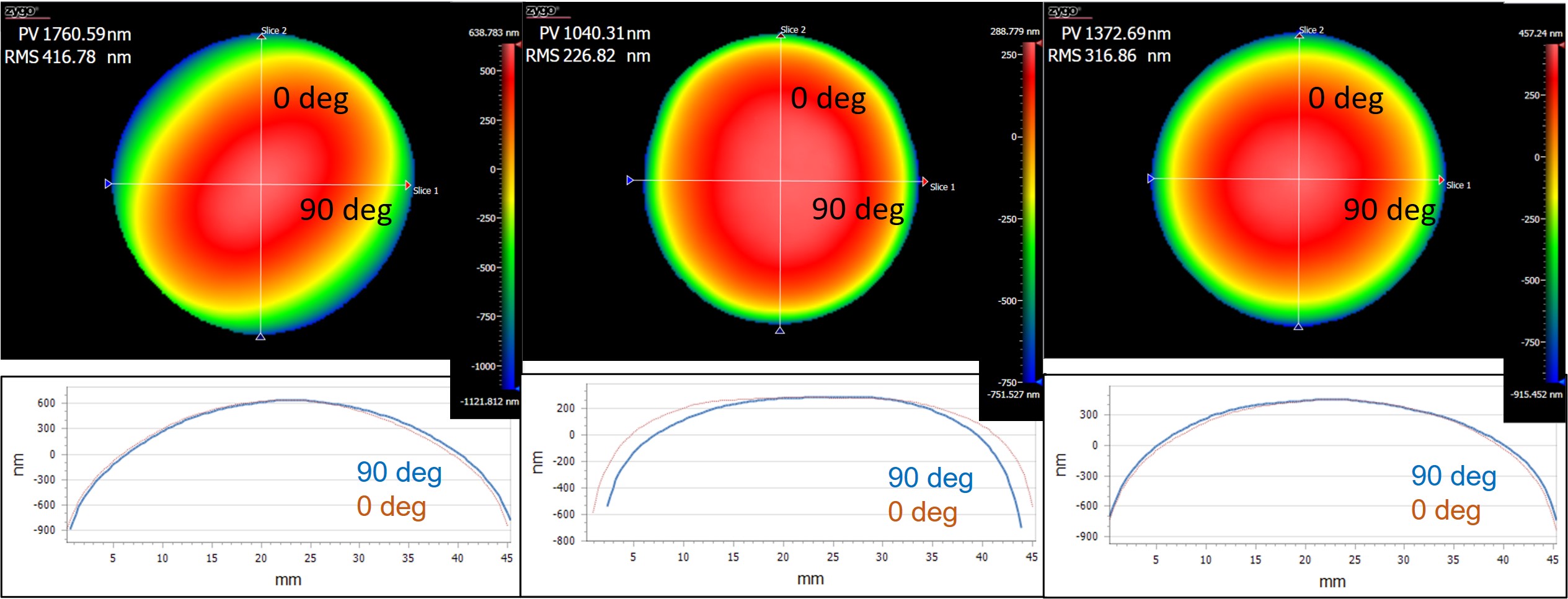}
    \caption{Surface form error map (tilt removed) of three Glassomer samples (01 to 03 from left to right) measured with the Zygo interferometer after polishing.}
    \label{fig:FSmet:form}
\end{figure}

Figure~\ref{fig:FSmet:form} presents the surface form error maps after polishing for each sample measured on the Zygo GPI-XP using a $\sim$\SI{45}{\mm} diameter aperture with tilt removed. Each sample retained a convex form, albeit with a significant reduction in PV, the line plots (Figure~\ref{fig:FSmet:form} \textit{lower}) correspond with the equivalent orientations in Figure~\ref{fig:FSmet:profil}. The PV (tilt removed) and PV (tilt, power, and astigmatism removed) are summarised in Table~\ref{tab:FS_met:summary} and contrasted against the as-printed results.   

{\renewcommand{\arraystretch}{1.25}
\begin{table}[h]
\centering
\caption{Fused silica: a collation of the initial (as-printed) and end (polished) metrology results for surface form and surface texture/micro-roughness.}
\begin{tabular}{|l||l|l|l|l||l|l|}
\hline
\multirow{3}{*}{Sample} & \multicolumn{4}{|c||}{Surface form error (PV)} & \multicolumn{2}{|c|}{Surface roughness (RMS)} \\
\cline{2-7}
& \multicolumn{2}{|c|}{As-printed} & \multicolumn{2}{|c||}{Polished} & As-printed& Polished\\
\cline{2-7}
& \ang{0} [\SI{}{\micro\metre}] & \ang{90} [\SI{}{\micro\metre}]& - (Z0 $\rightarrow$ Z2)* [\SI{}{\micro\metre}] & - (Z0 $\rightarrow$ Z6)* [\SI{}{\micro\metre}] & [\SI{}{\nm}] & [\SI{}{\nm}] \\
\hline
Glassomer \#1 & 131 & 105 & 1.8 & 0.4 & 1320 & 0.96 \\
Glassomer \#2 & 403 & 390 & 1.0 & 0.5 & 1270 & 0.97 \\
Glassomer \#3 & 270 & 226 & 1.4 & 0.5 & 1390 & 1.00 \\   
\hline
\multicolumn{7}{l}{* Zernike terms removed from the interferogram.}
\end{tabular}
\label{tab:FS_met:summary}
\end{table}}

\subsection{Silicon carbide}
\subsubsection{Micro-roughness after polishing}
\begin{figure}
    \centering
    \includegraphics[width = 0.95\textwidth]{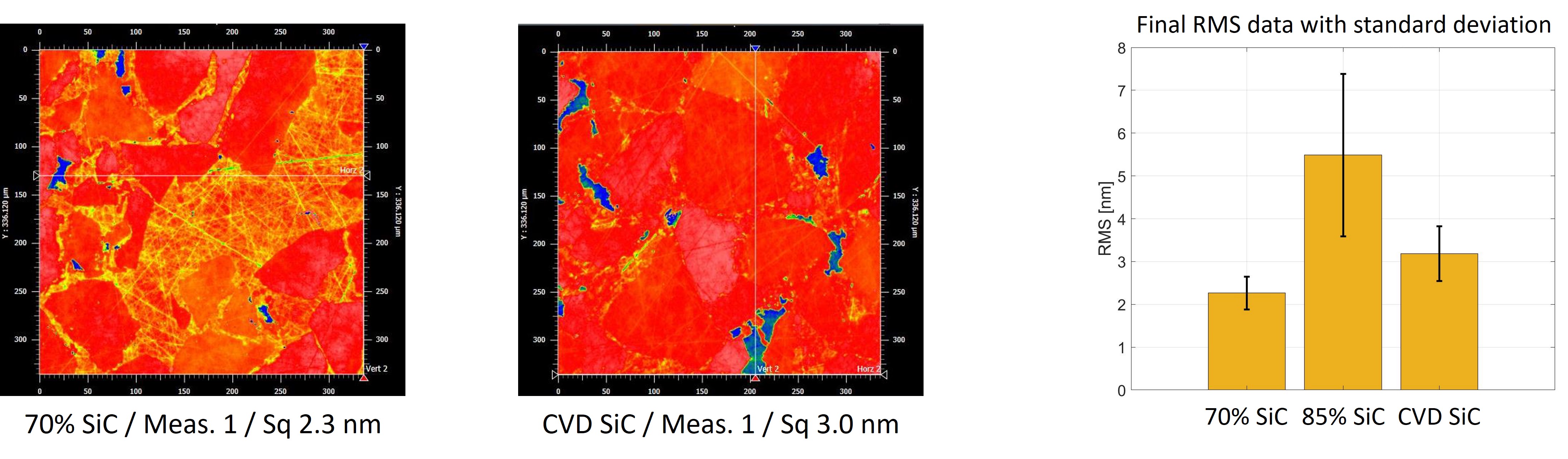}
    \caption{Micro-roughness of the SiC + Si samples: \textit{left} interferograms of the optical surfaces; and \textit{right} the average of five Sq measurements and the associated standard deviation.}
    \label{fig:SiCmet_micro}
\end{figure}

The micro-roughness was evaluated at five, evenly spaced, locations across a \SI{20}{\mm} line that was centred on the origin of the samples. Figure~\ref{fig:SiCmet_micro} \textit{left} and \textit{middle} provide representative interferograms of the reflective surfaces of the samples, and Figure~\ref{fig:SiCmet_micro} \textit{right} summarises the average Sq for each sample with its corresponding standard deviation. The average Sq values for 70\% SiC, 85\% SiC and CVD SiC are \SI{2.33}{\nm}, \SI{5.49}{\nm} and \SI{3.19}{\nm} respectively. The Sq value for 70\% SiC is consistent with the micro-roughness measured on the OPTICON prototype (Section~\ref{subsec:IACmet}). The measurements for 85\% SiC demonstrated a high degree of variability along the polished surface with the dominant contributor to roughness being the step height between the materials. In addition and likewise, the CVD SiC did not provide the anticipated uniform, single material, coating that was anticipated, as demonstrated by the difference between the blue and the red areas in Figure~\ref{fig:SiCmet_micro} \textit{middle}.

\subsubsection{Electron dispersive X-ray spectroscopy}
\begin{figure}
    \centering
    \includegraphics[width = 0.95\textwidth]{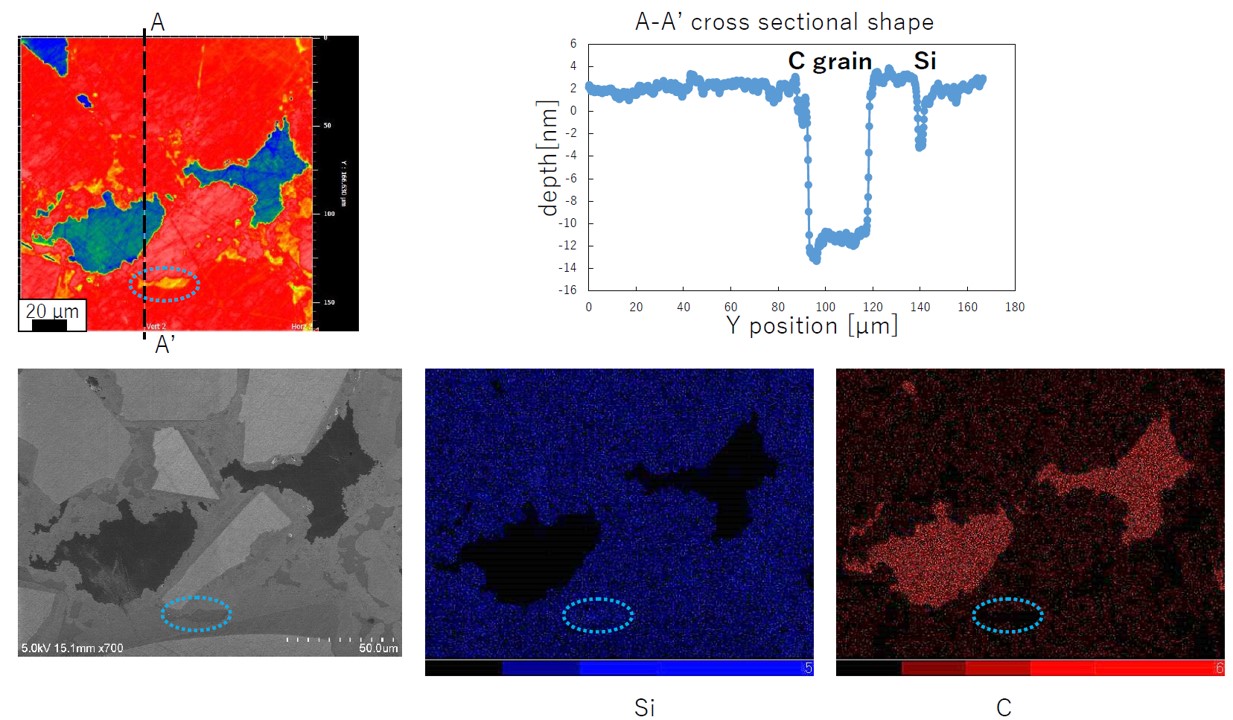}
    \caption{An EDX measurement and evaluation on the CVD SiC sample. The presence of C is clearly visible in the bottom right element map.}
    \label{fig:SiCmet_EDX}
\end{figure}

EDX measurements were performed on the 85\% SiC and the CVD SiC samples to define the elements present within the optical surface. In both surfaces, carbon (C) was present at a higher percentage than would be anticipated from the material and elemental composition information provided. The presence of C resulted in a step with a greater depth, due to its lower mechanical hardness (in this atomic arrangement) in comparison to SiC and Si. The presence of C in the CVD SiC coating, as shown in the EDX data in Figure~\ref{fig:SiCmet_EDX}, indicates that either C was present within the coating, or that the thickness of the coating was not sufficient for the lapping and polishing process and that the majority of the coating had been removed. 

\subsection{Discussion and conclusion}
A first trial at polishing AM fused silica samples yielded a micro-roughness values of $\le$\SI{1}{\nm} Sq, which clearly demonstrates the potential of this material to deliver mirrors for short wavelength applications. The Sq achieved within this study is not considered, at this time, the limit of the AM fused silica material: first, there are no equivalent data from the M.M. 8400 lapper/polisher for conventionally manufactured fused silica to indicate a micro-roughness limit for the combination of machine and material; and second, it was anticipated that moving to pitch polishing would have lowered the AM fused silica micro-roughness further, but this was outside the scope of this study.  

The study into different finishing processes highlighted the feasibility to achieve $<$\SI{6}{\nm} Sq on all samples and to achieve $\sim$\SI{2}{\nm} Sq on the 70\% SiC surface. Despite the challenge associated with the presence of C affecting the micro-roughness, these results demonstrate the potential for immediate applications in the infrared to visible. The lapping and polishing cycles between the OPTICON prototype and the disk samples, were undertaken on different machines and it is suspected that the machine used for the three samples was not capable of creating the same surface quality that was achieved on the OPTICON prototype.  

\section{SUMMARY \& FUTURE WORK}
\label{sec:future}

This paper has presented a proof-of-concept investigation into the use of commercially available AM SiC + Si and AM fused silica towards mirror fabrication for short wavelength applications (visible, UV and X-ray). Both materials, when lapped and polished, exhibited micro-roughness values with excellent potential and the materials require further investigation. The exploration of the AM SiC + Si samples for mirror fabrication followed a successful trial at polishing the OPTICON prototype (Section~\ref{sec:IAC}). The initial results are encouraging, but challenges with material compositions were encountered with the SiC + Si samples: steps between materials, and the elemental composition. Routes to mitigate these are described below. AM fused silica printed using vat photopolymerisation has the potential to enable new opportunities in low mass mirror fabrication, given the role of fused silica across astronomical hardware. The first fused silica results were very positive and the next steps are to understand what limited further improvement.       

To determine the limitation of the M.M. 8400 grinder/polisher to achieve sub-nanometre micro-roughness and to compare the AM fused silica against conventional fused silica, future work will lap and polish these materials simultaneously. The AM fused silica sample dimensions will be identical to this study and the samples have been purchased. 

The motivation for using AM fused silica is the increased design freedom that the print method offers, therefore a next generation of samples will be lapped and polished that exhibit low mass structures with two optical prescriptions: flat and convex. The lightweight samples are \SI{50}{\mm} in diameter, open back, and use a TPMS diamond lattice for the low mass structure, in addition, the convex design has a \SI{100}{\mm} spherical radius of curvature. Full details regarding the design of the lightweight samples can be found in \textit{Lister, et al., (2024)~\cite{Lister2024}}.  

Two approaches will be explored to advance SiC + Si for AM mirror fabrication. First, alternative processing steps prior to the CVD SiC coating will be explored to understand where in the optical fabrication chain is optimal to apply the CVD coating. The advantage of the CVD coating is that the optical surface is generated within a layer of a single material, which enables the use of fine-processing techniques, such as plasma polishing. Currently, with the presence of C within the material, this technique is not viable.

Second, due to the steps created at the interface between materials at the nanometre-scale which impact the surface roughness, routes mitigate the steps using optical coatings will be explored. Typically, sputtered optical coatings replicate the structure from beneath; however, given the nanometre-scale step height, the accuracy of the replication is unknown and may result in a smoothing at the micro-scale. Further, a coating methodology, such as planarization~\cite{Stolz2015}, which corrects for surface height defects could be applied to use the optical coating to compensate for the step difference observed.  

AM in ceramic materials, which are commonly associated with conventional low mass mirror fabrication, are commercially available, which allow the benefits of the AM design freedom, low mass geometries and part consolidation, to be combined with the materials capable of the low micro-roughness. This paper has outlined the potential of AM SiC + Si and AM fused silica to deliver low micro-roughness, a requirement of short wavelength applications, but the paper has also highlighted the need for further research to explore the AM ceramics material in further depth, and to find the most suitable optical fabrication chain to produce AM ceramic mirrors of high quality. 
 
\acknowledgments 
The authors acknowledge the UKRI Future Leaders Fellowship ‘Printing the future of space telescopes’ under grant \# MR/T042230/1 and the advice provided by Patrick Risch (Glassomer) and Markus Demharter (SGL Carbon). N. Yu acknowledges the Royal Society under grant \# IEC/R3/213107.

The SICAPRINT prototype presented in Section~\ref{sec:IAC} was printed as part of the European Union's Horizon 2020 research and development programme under grant agreement \# 730890 (OPTICON Work package 5). The design was created through a collaboration of Instituto de Astrof\'{i}sica de Canarias and TNO in the Netherlands. C. Atkins acknowledges the support of F. Tenegi-Sangin\'{e}s and A. Vega-Moreno in providing prototype information. C. Atkins acknowledges the role of N. Yu (co-author) in seeing the potential of the OPTICON prototype and in making the connection with Osaka University. 

\bibliography{report} 
\bibliographystyle{spiebib} 

\end{document}